\documentclass[fdp,a4paper,fleqn]{w-art}
\usepackage{times,cite,w-thm}
\usepackage{amsmath, amssymb, graphics, epsfig}
\usepackage{calc, color, sidecap, caption2}
\usepackage{multirow,bigdelim}
\theoremstyle{plain}

\theoremstyle{definition}

\usepackage[]{graphicx}
\begin{document}

\DOIsuffix{theDOIsuffix}
\Volume{55}
\Month{01}
\Year{2012}
\pagespan{1}{}
\keywords{Scattering Amplitudes, Perturbation Theory}



\title[Exploration of the Tree-Level S-Matrix of Massless Particles]{Exploration of the Tree-Level S-Matrix of Massless Particles}


\author[P. Benincasa]{Paolo Benincasa%
  \footnote{e-mail:~\textsf{paolo.benincasa@usc.es}.}}
\address{Departamento de F{\'i}sica de Part{\'i}culas, Universidade de Santiago de Compostela, E-157782, Santiago de
         Compostela, Spain}

\begin{abstract}
 In recent years, the BCFW construction provided a very powerful tool for computing scattering amplitudes as well as it shed
 light on the perturbation theory structure. In this talk, I discuss the long-standing issue of the boundary term arising
 when the amplitudes do not vanish as some momenta are taken to infinity along some complex direction. In particular, we 
 provide a new set of on-shell recursion relations valid for such theories and discuss its consequences on our understanding
 on the perturbation theory structure of the S-Matrix. 
\end{abstract}

\maketitle                   

\section{Introduction}

Interacting theories in asymptotically flat space time can be studied at weak coupling through the analysis of the
scattering amplitudes, which provide the probability that a certain number of asymptotic states scatter to produce
other asymptotic states. 

The best way that scattering amplitudes have been understood so far is by mean of the
Feynman diagram representation, which are diagrammatic rules coming from a Lagrangian formulation of the theories.
Such a representation makes manifest two basic properties of the theories: Poincar{\'e} invariance and the locality
of the interactions. There are, however, some side-effects. First of all, the individual Feynman diagrams may break other 
symmetries of the theories ({\it e.g.} it is indeed the case for gauge symmetries). Furthermore, the number of diagrams
dramatically increases with the increase of the number of external states, making the calculation more and more cumbersome.
Finally, the simplicity of the scattering amplitude can be hidden, as it happens for example in the case of scattering
of gluons and gravitons at tree level. In the former case, after  having summed up a high number of Feynman diagrams, the 
$n$-gluon amplitudes turns out to be zero if all the external gluons are in the same helicity state (or at most one gluon 
is in a different helicity state with respect to the others), while it acquire the simple Parke-Taylor form for MHV 
amplitudes \cite{Parke:1986gb}. In the latter case, instead, the computation of the $n$-graviton amplitude would require to
keep into account all the $k$-point vertices (with $k\,\le\,n$).

It is therefore fair to ask whether it is possible to formulate some other representation which can point out the 
simplicity of the scattering amplitudes, and to which extent we really understand perturbation theory.

The question about the existence of other diagrammatic representations can be already answered positively. A first example
is the CSW-expansion \cite{Cachazo:2004kj}, which is characterised by off-shell diagrams constructed out of just 
MHV-vertices and makes manifest locality while the Lorentz invariance of the theory is broken by the individual diagrams.
This representation is however not general and it holds just for Yang-Mills theories. A second (and more general)
example is provided by the BCFW-construction \cite{Britto:2004aa, Britto:2005aa}, which is instead an on-shell diagrammatic 
expansion in which the gauge invariance of the theory is not broken at intermediate stages, Lorentz invariance stays 
manifest. The price one pays is that the individual on-shell diagrams breaks locality. 

It is instead probably needed a deeper understanding of the perturbative structure of interacting theories, even at tree
level. First of all, one can try to use Occam's razor and try to understand which is the minimal set of assumptions to
formulate a general S-matrix theory in flat space. In particular, the BCFW-construction seems to suggest that such a minimal
set can be formed by Poincar{\'e} invariance, analyticity, Existence of one-particle states, Locality of the whole S-matrix
\cite{Benincasa:2007xk}. This fact, together with its intrinsic generality as a method, suggests the BCFW-construction
as a good starting problem to concretely approach this problem.

\section{BCFW-construction}\label{BCFW}

A way to understand the structure of scattering amplitudes is through the analysis of their singularities. This can 
however be a really difficult problem to face, given that the scattering amplitudes can be seen as analytic functions of
Lorentz invariants and the number of Lorentz invariants increases with the number of external states. A drastic 
simplification can be obtained by introducing a $1$-parameter deformation of the complexified momentum space, such
that the both the massless on-shell condition $p^2\,=\,0$ and the momentum conservation are preserved \cite{Britto:2005aa}. 
A deformation with such characteristics is indeed not unique, and the simplest one is obtained by deforming the momenta
of two particles (namely labelled by $i$ and $j$), leaving the ones of the others unchanged:
\begin{equation}\label{BCFWdef}
  p^{\mbox{\tiny $(i)$}}(z)\,=\,p^{(\mbox{\tiny $j$})}-zq,\qquad
  p^{\mbox{\tiny $(j)$}}(z)\,=\,p^{(\mbox{\tiny $j$})}+zq,\qquad
  p^{\mbox{\tiny $(k)$}}(z)\,=\,p^{\mbox{\tiny $(k)$}},\:\forall\,k\neq\,\{i,\,j\},
\end{equation}
with the on-shell condition fixing the momentum $q$ to satisfy the following relations:
\begin{equation}\label{BCFWq}
 q^2\:=\:0,\qquad  p^{\mbox{\tiny $(i)$}}\cdot q\:=\:0\:=\: p^{\mbox{\tiny $(j)$}}\cdot q.
\end{equation}
With such a deformation, the amplitudes are mapped into a $1$-parameter family of amplitudes 
$M_{\mbox{\tiny $n$}}\,\rightarrow\,M_{\mbox{\tiny $n$}}^{\mbox{\tiny $(i,j)$}}(z)$. One can therefore analyse the 
singularity structure of the amplitudes as a function of $z$ only. Generally speaking, the scattering amplitudes are 
characterised by both poles and branch points. Focusing on the pole structure is however equivalent to focusing on the
tree level and the poles are provided by the $z$-dependent internal propagators, {\it i.e.} they are present in those
channels where the internal momentum can be written as a sum containing either $p^{\mbox{\tiny $(i)$}}$ or 
$p^{\mbox{\tiny $(j)$}}$
\begin{equation}\label{BCFWpoles}
 \frac{1}{[P^2_{\mbox{\tiny $\mathcal{I}_k$}}(z)]^2}\:=\:\frac{1}{P_{\mbox{\tiny $\mathcal{I}_k$}}^2-
  2z\left(P_{\mbox{\tiny $\mathcal{I}_k$}}\cdot q\right)}
 \quad\Rightarrow\quad
 z_{\mbox{\tiny $\mathcal{I}_k$}}\:=\:\frac{P_{\mbox{\tiny $\mathcal{I}_k$}}^2}{P_{\mbox{\tiny $\mathcal{I}_k$}}\cdot q}.
\end{equation}
As the location of a pole is approached, the internal propagator goes on-shell, and the amplitude factorises, with
the residue of the pole which is given in terms of the product of two on-shell sub-amplitudes
\begin{equation}\label{BCFWfact}
 M_{n}^{\mbox{\tiny $(i,j)$}}(z)\:\overset{\mbox{\tiny $z\rightarrow z_{\mbox{\tiny $k$}}$}}{\sim}\:
  \frac{M_{\mbox{\tiny L}}^{\mbox{\tiny $(i,j)$}}(z_{\mathcal{I}_k})
        M_{\mbox{\tiny R}}^{\mbox{\tiny $(i,j)$}}(z_{\mathcal{I}_k})}{
  P^2_{\mbox{\tiny $\mathcal{I}_k$}}(z)}.
\end{equation}
This suggests the possibility to connect the whole amplitude to the on-shell lower-point amplitudes
\begin{equation}\label{BCFWres}
 0\:=\:\frac{1}{2\pi i}\oint_{\mbox{\tiny $\mathcal{R}$}}\frac{dz}{z}M_{n}^{\mbox{\tiny $(i,j)$}}(z)\:=
      \:M_{n}^{\mbox{\tiny $(i,j)$}}(0)-\sum_{k\in\mathcal{P}^{\mbox{\tiny $(i,j)$}}}
     \frac{M_{\mbox{\tiny L}}^{\mbox{\tiny $(i,j)$}}(z_k)M_{\mbox{\tiny R}}^{\mbox{\tiny $(i,j)$}}(z_k)}{P^2_k}
     -\mathcal{C}_{n}^{\mbox{\tiny $(i,j)$}},
\end{equation}
where the integration is performed along the whole Riemann sphere $\mathcal{R}$, $M_{n}^{\mbox{\tiny $(i,j)$}}(0)$ coincides
with the physical amplitude and the boundary term $\mathcal{C}_{n}^{\mbox{\tiny $(i,j)$}}$ is the residue of the singularity
at infinity, which is zero if the amplitude vanishes as $z\,\rightarrow\infty$ (constructibility condition). In such a case, 
\eqref{BCFWres} implies that the amplitude can be written as sum of products of on-shell lower-points amplitudes
\begin{equation}\label{BCFWrr}
  M_n\:=\:\sum_{k\in\mathcal{P}^{\mbox{\tiny $(i,j)$}}}
   \frac{M_{\mbox{\tiny L}}^{\mbox{\tiny $(i,j)$}}(\hat{i},\mathcal{I}_k,-\hat{P}_{\mbox{\tiny $i\mathcal{I}_k$}})
   M_{\mbox{\tiny R}}^{\mbox{\tiny $(i,j)$}}(\hat{P}_{\mbox{\tiny $i\mathcal{I}_k$}}, \mathcal{J}_{k}, \hat{j})}{
   P_{\mbox{\tiny $i\mathcal{I}_k$}}^2}.
\end{equation}
This structure has been shown to hold in Yang-Mills theory \cite{Britto:2005aa}, in GR \cite{Benincasa:2007qj}, as well
as for $\mathcal{N}=4$ Supersymmetric Yang-Mills theory and $\mathcal{N}=8$ Supergravity \cite{ArkaniHamed:2008gz}.

\section{BCFW construction: generalised recursive relations}\label{Bound}

There are however several theories ({\it e.g.:} Q.E.D., Einstein-Maxwell) which do not satisfy the constructibility 
condition and, therefore, the boundary term is non-zero. In this case, the $1$-parameter family of amplitudes 
$M_{n}^{\mbox{\tiny $(i,j)$}}(z)$ acquires the form
\begin{equation}\label{BoundRes}
   M_{n}^{\mbox{\tiny $(i,j)$}}(z)\:=\:\sum_{k\in\mathcal{P}^{\mbox{\tiny $(i,j)$}}}
    \frac{M_{\mbox{\tiny L}}^{\mbox{\tiny $(i,j)$}}(z_k)M_{\mbox{\tiny R}}^{\mbox{\tiny $(i,j)$}}(z_k)}{P^2_k(z)}
    +\mathcal{C}_{n}^{\mbox{\tiny $(i,j)$}}(z).
\end{equation}
Since the first term in \eqref{BoundRes} contains all the poles at finite location, the function
$\mathcal{C}_{n}^{\mbox{\tiny $(i,j)$}}(z)$ is just a polynomial in $z$ of order $\nu$, 
\begin{equation}\label{Bound0}
 \mathcal{C}_{n}^{\mbox{\tiny $(i,j)$}}(z)\:=\:\mathcal{C}_{n}^{\mbox{\tiny $(i,j)$}}+
    \sum_{l=1}^{\nu}a_{l}^{\mbox{\tiny $(i,j)$}}z^l,
\end{equation}
where $\nu$ is as well the order the amplitude $ M_{n}^{\mbox{\tiny $(i,j)$}}(z)$ diverges with, as $z$ is taken to 
infinity, and the $0$-th order term is the only which survives in \eqref{BCFWres} contributing to the physical amplitude.
The question we now need to answer is how to actually compute the boundary term\footnote{Some steps in this direction has 
been done in specific cases in \cite{Feng:2009ei, Feng:2010ku}}. The knowledge of the poles does not seem to be enough
to fully determine the amplitude when the constructibility condition is not satisfied. This means that we need to resort
to some new quantity. Given that a tree-level amplitude is just a rational function, the natural quantities to consider
together with poles are the zeroes \cite{Benincasa:2011kn}. Let $\{z_0^{\mbox{\tiny $(s)$}}\}$ be subset of zeroes with 
multiplicity $m^{\mbox{\tiny $(s)$}}$ and $\gamma_{\mbox{\tiny $0$}}^{\mbox{\tiny $(s)$}}$ be a contour including just
$z_0^{\mbox{\tiny $(s)$}}$ and no other zero or poles. From the expressions \eqref{BoundRes} and \eqref{Bound0}, one
gets the following system of equations
\begin{equation}\label{BoundEqs}
  \begin{split}
   0\:=\:\frac{1}{2\pi i}\oint_{\mbox{\tiny $\gamma_0^{(s)}$}}dz\:
   &\frac{M_{n}^{\mbox{\tiny $(i,j)$}}(z)}{\left(z-z_0^{\mbox{\tiny $(s)$}}\right)^{r}}\:=\:
   \left(-1\right)^{r-1}\sum_{k=1}^{N_{\mbox{\tiny $P$}}^{\mbox{\tiny fin}}}
   \frac{M_{\mbox{\tiny L}}^{\mbox{\tiny $(i,j)$}}(z_k)M_{\mbox{\tiny R}}^{\mbox{\tiny $(i,j)$}}(z_k)}{
   \left(-2P_k\cdot q\right)\left(z_0^{\mbox{\tiny $(s)$}}-z_k\right)^{r}}+
   \delta_{\mbox{\tiny $r,1$}}\mathcal{C}_{\mbox{\tiny $n$}}^{\mbox{\tiny $(i,j)$}}+\\
   &\hspace{.2cm}+
   \sum_{l=1}^{\nu}\frac{l!}{\left(l-r+1\right)!\left(r-1\right)!}a_{l}^{\mbox{\tiny $(i,j)$}}z_{\mbox{\tiny $0$}}^{l-r+1},
   \quad \mbox{with }
   \left\{
    \begin{array}{l}
     r\,=\,1,\ldots,m^{\mbox{\tiny $(s)$}}\\
     s\,=\,1,\ldots,n_z
    \end{array}
   \right. ,
  \end{split}
\end{equation}
whose solution reveals a connection between $\mathcal{C}_{\mbox{\tiny $n$}}^{\mbox{\tiny $(i,j)$}}$ and a sum of products
of on-shell scattering amplitudes with fewer external states. The explicit expression itself (see \cite{Benincasa:2011kn}) 
is not particularly illuminating. However, once it is reinserted  in \eqref{BoundRes}, it allows to rewrite the 
scattering amplitude in such a way that the overall structure of the BCFW expansion is still preserved 
\cite{Benincasa:2011kn}
\begin{equation}\label{GenBCFW}
M_n\:=\:\sum_{k\in\mathcal{P}^{\mbox{\tiny $(i,j)$}}}
   M_{\mbox{\tiny L}}^{\mbox{\tiny $(i,j)$}}(\hat{i},\mathcal{I}_k,-\hat{P}_{\mbox{\tiny $i\mathcal{I}_k$}})
   \frac{f_{\mbox{\tiny $i\mathcal{I}_k$}}^{\mbox{\tiny $(\nu,n)$}}}{P_{\mbox{\tiny $i\mathcal{I}_k$}}^2}
   M_{\mbox{\tiny R}}^{\mbox{\tiny $(i,j)$}}(\hat{P}_{\mbox{\tiny $i\mathcal{I}_k$}}, \mathcal{J}_{k}, \hat{j}),
\end{equation}
with the ``weights'' $f_{\mbox{\tiny $i\mathcal{I}_k$}}^{\mbox{\tiny $(\nu,n)$}}$ being
\begin{equation}\label{GenWeights}
  f_{\mbox{\tiny $i\mathcal{I}_k$}}^{\mbox{\tiny $(\nu,n)$}}\:=\:
  \left\{
   \begin{array}{l}
    1,\hspace{4cm} \nu\,<\,0,\\
    \phantom{\ldots}\\
    \prod_{l=1}^{\nu+1}\left(1-\frac{P_{\mbox{\tiny $i\mathcal{I}_k$}}^2}{P_{\mbox{\tiny $i\mathcal{I}_k$}}^2
     \left(z_0^{\mbox{\tiny $(l)$}}\right)}\right),\quad \nu\,\ge\,0,
   \end{array}
  \right.
\end{equation}
The recursion relation \eqref{GenBCFW} allows to state that the BCFW-structure is generalised to any consistent theory
at tree level, meaning that, for any consistent theory, the amplitudes can be expressed in terms of sum of products of
lower-point on-shell amplitudes and propagators, now with a simple weight \eqref{GenWeights} which depends on the location 
of a subset of zeroes of the amplitudes. Iterating the recursive relation, one can express an $n$-point amplitudes just
in terms of propagators and the smallest amplitude present in the theory. For theories with $3$-particle interactions,
this is indeed the $3$-particle amplitude, which is determined by momentum conservation and Lorentz invariance
\cite{Benincasa:2007xk}. In the case instead the smallest interaction is an higher point one, generally speaking one
can always introduce an auxiliary massive particles to define $3$-particle interactions and than integrate it out,
as it has been done in the case of $\lambda\phi^4$ \cite{Benincasa:2007xk}. Even if this might not be computationally
convenient, it points out that also these theories can be determined by $3$-particle amplitudes. 

\section{Zeroes and collinear/multiparticle limits}\label{ZeroLim}

If on one side \eqref{GenBCFW} is a valid mathematical expression which reveals a general structure for tree-level 
amplitudes, on the other side it seems to be not really practical given that a general way to determine the zeroes is not
known. However, the recursion relation \eqref{GenBCFW} provides a representation for the amplitudes and, therefore, must
factorise properly when collinear/multiparticle limits are taken. Such limits can be grouped in four classes
\begin{equation}\label{ZeroLimCl}
 \begin{split}
  &\lim_{\mbox{\tiny $P_{i\mathcal{I}_k}^2$}\:\rightarrow\:0}
    P_{\mbox{\tiny $i\mathcal{I}_k$}}^2\,M_n\:=\:
   M(i,\,\mathcal{I}_k,\,-P_{i\mathcal{I}_k})\, M(P_{i\mathcal{I}_k}, \mathcal{J}_k, j),\\
  &\lim_{\mbox{\tiny $P_{\mathcal{K}}^2$}\:\rightarrow\:0}
    P_{\mbox{\tiny $\mathcal{K}$}}^2\,M_n\:=\:
   M_{s+1}(\mathcal{K},\,-P_{\mathcal{K}})\, M_{n-s+1}(P_{\mathcal{K}}, \mathcal{Q}, i, j),\\
  &\lim_{\mbox{\tiny $P_{k_1 k_2}^2$}\:\rightarrow\:0}
    P_{\mbox{\tiny $k_1 k_2$}}^2\,M_n\:=\:
   M_3(k_1,\,k_2,\,-P_{k_1 k_2})\, M_{n-1}(P_{k_1 k_2}, \mathcal{K}, i, j),\\
  &\lim_{\mbox{\tiny $P_{ij}^2$}\:\rightarrow\:0}
    P_{\mbox{\tiny $ij$}}^2\,M_n\:=\:
   M_{3}(i,\,j,\,-P_{ij})\, M_{n-1}(P_{ij}, \mathcal{K}),
 \end{split}
\end{equation}
and their analysis leads to the following conditions of the zeroes
\begin{equation}\label{ZerosCond}
 \begin{split}
     &\hspace{2.5cm}P_{ik}^2(z_0^{\mbox{\tiny $(l)$}})\:=\:\langle i,k\rangle\alpha_{ik}^{\mbox{\tiny $(l)$}}[i,j],\qquad
      P_{jk}^2(z_0^{\mbox{\tiny $(l)$}})\:=\:\langle i,j\rangle\alpha_{jk}^{\mbox{\tiny $(l)$}}[j,k],\\
     &\hspace{2.5cm}\lim_{P_{\mathcal{K}}^2\rightarrow0}f_{\mbox{\tiny $i\mathcal{I}_k$}}^{\mbox{\tiny $(\nu,\,n)$}}\:=\:
      f_{\mbox{\tiny $i\mathcal{I}_k$}}^{\mbox{\tiny $(\nu,\,n-s+1)$}},\qquad
      \lim_{P_{i\mathcal{I}_k}^2\rightarrow0}f_{i\mathcal{I}_k}^{\mbox{\tiny $(\nu,n)$}}\,=\,1,\\
     &\hspace{2.5cm}\lim_{[k_1,k_2]\rightarrow0}f_{\mbox{\tiny $i\bar{k}$}}^{\mbox{\tiny $(\nu,n)$}}\:=\:
       f_{\mbox{\tiny $i(k_1 k_2)$}}^{\mbox{\tiny $(\nu,n-1)$}},\qquad
      \lim_{\langle k_1,k_2\rangle\rightarrow0}f_{\mbox{\tiny $j\bar{k}$}}^{\mbox{\tiny $(\nu,n)$}}\:=\:
       f_{\mbox{\tiny $j(k_1 k_2)$}}^{\mbox{\tiny $(\nu,n-1)$}},\\
     &\lim_{[i,j]\rightarrow0}\sum_{k}(-1)^{2(h_i+h_j+h_k)+\delta+\nu+1}
      \left[
       \left(\frac{\langle i,k\rangle}{\langle i,j\rangle}\right)^{\delta-1}
       \left(\frac{[i,j]}{[i,k]}\right)^{2h_i+\delta-\nu}
       \frac{\mathcal{H}_{n-1}^{(k)}}{\prod_{l=1}^{\nu+1}\alpha_{ik}^{\mbox{\tiny $(l)$}}}\right]\:=\:1,\\
     &\lim_{\langle i,j\rangle\rightarrow0}\sum_k(-1)^{2(h_i+h_k)+\delta+\nu+1}
      \left[
       \left(\frac{[j,k]}{[i,j]}\right)^{\delta-1}
       \left(\frac{\langle i,j\rangle}{\langle j,k\rangle}\right)^{\delta-2h_j-\nu}
       \frac{\tilde{\mathcal{H}}_{n-1}^{(k)}}{\prod_{l=1}^{\nu+1}\alpha_{jk}^{\mbox{\tiny $(l)$}}}
      \right]\:=\:1,
    \end{split}
\end{equation}
where the notation has been detailed in \cite{Benincasa:2011kn}. Here, it is important to know just that 
$\mathcal{H}$ is a dimensionless helicity factor, $\delta$ is the number of derivatives of the $3$-particle interactions,
and that the BCFW-deformation \eqref{BCFWdef} has been implemented by considering the bispinorial representation of the
momenta $p_{a\dot{a}}\,=\,\lambda_{a}\tilde{\lambda}_{\dot{a}}$ and shifting the spinors as 
$\tilde{\lambda}^{\mbox{\tiny $(i)$}}(z)\,=\,\tilde{\lambda}^{\mbox{\tiny $(i)$}}-z\tilde{\lambda}^{\mbox{\tiny $(j)$}}$,
$\lambda^{\mbox{\tiny $(j)$}}\,=\,\lambda^{\mbox{\tiny $(j)$}}+z\lambda^{\mbox{\tiny $(i)$}}$. The relations 
\eqref{ZerosCond} suggests the possibility to connect the ``weights'' of the $n$-particle amplitudes with the ones of 
the lower point ones. At present, such a general connection is still missing. However, the conditions \eqref{ZerosCond}
are solvable for a number of cases\footnote{A further analysis of the zeroes was recently done in \cite{Feng:2011jxa}.}.

\section{Soft limits of $3$-particle amplitudes and complex-UV behaviour}\label{Soft}

The analysis in the previous section points out that the collinear limit $P_{ij}^2\,\rightarrow\,0$ appears
in the BCFW-representation as a soft singularity \cite{Schuster:2008nh}, and it can be taken as 
$\hat{p}^{\mbox{\tiny $(i)$}}\,\rightarrow\,0$ or $\hat{p}^{\mbox{\tiny $(j)$}}\,\rightarrow\,0$. 
Furthermore, in this limit the only on-shell diagrams which contribute
are the ones showing a three-particle amplitude with one of the deformed momenta, that is the momentum which becomes soft.
If the softening of this momentum is able to produce the correct pole than the amplitude factorises properly in this channel
and the standard BCFW-representation is valid, otherwise are the weights which have to contribute to the formation of the
right singularity, as implied by the first line in \eqref{ZerosCond}. The soft behaviour of the $3$-particle amplitudes
can therefore be taken as a criterion for the validity of the standard BCFW-representation. The analysis of the
collinear limit $P_{ij}^2\,\rightarrow\,0$ on the generalised BCFW-representation \eqref{GenBCFW} allows to prove that
such a soft-behaviour coincides with the large-$z$ behaviour of the whole amplitude \cite{Benincasa:2011pg}. A rigorous 
proof of this equality in the case of the standard BCFW-representation is not known, but it has been checked for all the 
known cases. Such a behaviour can be generally written just in terms of the number of the derivatives of the $3$-particle
interaction and the helicity of the deformed particles
\begin{equation}\label{SoftBehav}
 \nu\:=\:\delta+2h_i \qquad\mbox{ and/or }\qquad \nu\:=\:\delta-2h_j,
\end{equation}
which are respectively valid if the collinear limit $P_{ij}^2\,\rightarrow\,0$ translates to the soft limit 
$\hat{p}^{\mbox{\tiny $(i)$}}\,\rightarrow\,0$ or to $\hat{p}^{\mbox{\tiny $(j)$}}\,\rightarrow\,0$, while in the case
both these soft limits cane be taken, the two relations in \eqref{SoftBehav} coincide.
This strikingly implies that also the information about the complex-UV behaviour of the $n$-particle amplitude of
a given theory is encoded in their building blocks and it is independent on the number of external states.

\section{Exploration of the space of theories}\label{Expl}

It has been noticed in \cite{Benincasa:2011pg} that the conditions \eqref{ZerosCond} drastically simplifies in the
case of the $4$-particle amplitudes
\begin{equation}\label{ZerosCond4}
 \prod_{r=1}^{N_{P}^{\mbox{\tiny fin}}}P_{i s_r}^2(z_0^{\mbox{\tiny $(l)$}})\:=\:
     (-1)^{N_{P}^{\mbox{\tiny fin}}}\left(P_{ij}^2\right)^{N_{P}^{\mbox{\tiny fin}}},
\end{equation}
where $N_{P}^{\mbox{\tiny fin}}$ is the number of poles at finite location, which can be just one or two. Such a condition
allows to generalise the four-particle consistency test proposed in \cite{Benincasa:2007xk}. The idea is to compute the
$4$-particle amplitude through two different BCFW-deformations. Imposing that the two results coincide brings on non-trivial
constraints on the S-matrix \cite{Benincasa:2007xk}
\begin{equation}\label{FourPT}
 M_4^{\mbox{\tiny $(i,j)$}}(0)\:=\:M_4^{\mbox{\tiny $(i,k)$}}(0).
\end{equation}
In \cite{Benincasa:2007xk} this test allowed to rediscover the Jacobi-identity in Yang-Mills and $\mathcal{N}=1$ 
Supergravity, in which both the gauge symmetry and supersymmetry emerge from the consistency of the theory. The
generalised BCFW-representation, together with the condition on the zeroes \eqref{ZerosCond4}, allows to rediscover
the existent theories not satisfying the constructibility condition, which turn out to belong to the class of theories
characterised by the $3$-particle amplitudes with $(\mp s, \mp s, \pm s)$ and $(\mp s', \pm s', \mp s)$ 
(see table \ref{tab:SumS}).
 \begin{table}[t] 
        \centering
        \begin{tabular}{| l | l | l |}\hline
         $s$ & Conditions & Interactions\\\hline
         $s=0$ & $s'=\frac{1}{2}$, $\kappa=\kappa'$ & Yukawa\\\hline
         \multirow{3}{*}{$s=1$} &  $s'=0$, $\kappa=0$ & scalar QED and YM + scalars\\
         &  $s'=\frac{1}{2}$, $\kappa=0$ & QED and YM + fermions\\
         &  $s'=1$, $\kappa=\kappa'$ & YM\\\hline
         \multirow{3}{*}{$s=2$} &  $s'=0$, $\kappa=\kappa'$ & scalar GR\\
         &  $s'=\frac{1}{2}$, $\kappa=\kappa'$ & Fermion Gravity\\
         &  $s'=1$, $\kappa=\kappa'$ & Einstein-Maxwell\\
         &  $s'=\frac{3}{2}$, $\kappa=\kappa'$ & ${\cal N}=1$ supergravity\\
         &  $s'=2$, $\kappa=\kappa'$ & GR\\\hline
        \end{tabular}
       \caption{Summary of the theories characterised by couplings with $s$-derivative interactions.}
       \label{tab:SumS}
       \end{table}

The theories can be generically classified through the dimension of the $3$-particle coupling constant, or equivalently 
through the number of derivative of the $3$-particle interaction, which implies that $3$-particle amplitudes are 
characterised by well-defined helicity configurations. The consistency test seems to reveal the appearance of unexpected 
interactions in which the spin of the particles is higher than $2$. Some comments are now in order. The $4$-particle 
amplitudes in these theories are all characterised by the presence of just one factorisation channel. This means that the 
representation \eqref{GenBCFW} is still meaningful if and only if the deformation chosen has such a channel as a 
BCFW-channel. If it were otherwise, the $1$-parameter family of amplitudes would not have any pole at finite location and 
the whole amplitude would be given by the boundary term. Furthermore, in the former case, the absence of a second 
factorisation channel does not allow us to perform the analysis which brought to the relation \eqref{ZerosCond4} which fixes
the condition on the zeroes.
If one thinks about the limits in which the amplitudes become trivial as a generic property, it is reasonable to assume
that the condition \eqref{ZerosCond4} can still hold. This is indeed a strong assumption which needs to be checked.

Another characteristic of these theories is that the boundary term turns out to be a polynomial in the Lorentz invariants.
A simple dimensional analysis reveals that it has a structure of a contact interaction with 
$L_n\,=\,(\delta-2)n-2(\delta-3)$ derivatives which increase as $n$ increases. Therefore, these theories seem to be endowed 
with higher-derivative interaction terms with the increase of the number of external states. This can be interpreted as a 
signature of non-locality for these theories.

Furthermore, the consistency of the $4$-particle amplitudes for these theories indeed does not imply the consistency
neither of the whole theories nor of their tree-level. Pathologies can, for example, already arise at $5$-particle level:
the fact that the consistency test has been passed at $4$-particle level does not imply that it has to be passed 
generically. Furthermore, ghosts might appear at loops.

Finally, as a general comment, it is fair to point out that our analysis concerns just those theories whose propagators
are given by $1/P^2$.


\begin{acknowledgement}
 I would like to thank the organisers for the stimulating environment created at the XVII${}^{\mbox{\tiny th}}$ 
 European Workshop on String Theory. It is also a pleasure to thank Eduardo Conde who collaborated with me on the
 works which this talk is based on. My work is funded in part by MICINN under grant FPA2008-01838, by the Spanish 
Consolider-Ingenio 2010 Programme CPAN (CSD2007-00042) and by Xunta de Galicia (Conseller{\'i}a de Educac{\'i}on, grant 
INCITE09 206 121 PR and grant PGIDIT10PXIB206075PR) and by FEDER. I am supported as well by the MInisterio de Ciencia e 
INNovaci{\'on} through the Juan de la Cierva program.
\end{acknowledgement}

\bibliographystyle{fdp}
\bibliography{amplitudesrefs}

\providecommand{\WileyBibTextsc}{}
\let\textsc\WileyBibTextsc
\providecommand{\othercit}{}
\providecommand{\jr}[1]{#1}
\providecommand{\etal}{~et~al.}


\begin{thebibliography}{[10]}

\bibitem{Parke:1986gb}
 \textsc{S.\,J. Parke} and  \textsc{T.\,R. Taylor},
 \jr{Phys. Rev. Lett.} \textbf{56}, 2459 (1986).


\bibitem{Cachazo:2004kj}
 \textsc{F.~Cachazo},  \textsc{P.~Svrcek},  and  \textsc{E.~Witten},
 \jr{JHEP} \textbf{09}, 006 (2004).


\bibitem{Britto:2004aa}
 \textsc{R.~Britto},  \textsc{F.~Cachazo},  and  \textsc{B.~Feng},
 \jr{Nucl. Phys. B} \textbf{715}, 499 (2005).


\bibitem{Britto:2005aa}
 \textsc{R.~Britto},  \textsc{F.~Cachazo},  \textsc{B.~Feng},  and
  \textsc{E.~Witten},
 \jr{Phys. Rev. Lett.} \textbf{94} (2005).


\bibitem{Benincasa:2007xk}
 \textsc{P.~Benincasa} and  \textsc{F.~Cachazo}(2007).


\bibitem{Benincasa:2007qj}
 \textsc{P.~Benincasa},  \textsc{C.~Boucher-Veronneau},  and
  \textsc{F.~Cachazo},
 \jr{JHEP} \textbf{11}, 057 (2007).


\bibitem{ArkaniHamed:2008gz}
 \textsc{N.~Arkani-Hamed},  \textsc{F.~Cachazo},  and
  \textsc{J.~Kaplan},
 \jr{JHEP} \textbf{1009}, 016 (2010).


\bibitem{Feng:2009ei}
 \textsc{B.~Feng},  \textsc{J.~Wang},  \textsc{Y.~Wang},  and
  \textsc{Z.~Zhang},
 \jr{JHEP} \textbf{01}, 019 (2010).


\bibitem{Feng:2010ku}
 \textsc{B.~Feng} and  \textsc{C.\,Y. Liu},
 \jr{JHEP} \textbf{07}, 093 (2010).


\bibitem{Benincasa:2011kn}
 \textsc{P.~Benincasa} and  \textsc{E.~Conde},
 \jr{JHEP} \textbf{1111}, 074 (2011).


\bibitem{Feng:2011jxa}
 \textsc{B.~Feng},  \textsc{Y.~Jia},  \textsc{H.~Luo},  and
  \textsc{M.~Luo}(2011).


\bibitem{Schuster:2008nh}
 \textsc{P.\,C. Schuster} and  \textsc{N.~Toro},
 \jr{JHEP} \textbf{0906}, 079 (2009).


\bibitem{Benincasa:2011pg}
 \textsc{P.~Benincasa} and  \textsc{E.~Conde}(2011).


\end{thebibliography}

\end{document}